\documentclass[preprint,aps,showpacs,tightenlines]{revtex4}
\begin{document}
\title{Quantum Disentanglement \\in Long-range Orders and Spontaneous
Symmetry Breaking}
\author{Yu Shi}
\email[Email:]{ys219@cam.ac.uk}
\affiliation{Theory of Condensed
Matter, Cavendish Laboratory, University of Cambridge, Cambridge
CB3 0HE, United Kingdom}
\affiliation{Department of Applied
Mathematics and Theoretical Physics, University of Cambridge,
Wilberforce Road, Cambridge CB3 0WA, United Kingdom}

\begin{abstract}
We investigate the nature of quantum entanglement in long-range
orders and spontaneous symmetry breaking. It is shown that
diminishing of entanglement between the condensate mode and the
rest of the system underlies off-diagonal long-range order, which
is the hallmark of superconductivity and Bose-Einstein
condensation. It is also revealed that disentanglement underlies
various cases of long-range order and spontaneous symmetry
breaking. In the course of the discussion, we also present some
ideas on characterizing entanglement in many-body systems.
Especially, it is shown how the connected correlation functions
can be used in characterizing entanglements in a pure state.

{\bf Key words}: disentanglement, entanglement, long-range order,
spontaneous symmetry breaking

\end{abstract}

\pacs{PACS: 74.20.-z, 75.10.-b, 05.30.-d, 03.65.-w}

\maketitle

\section*{1. Introduction}

Long-range order and spontaneous symmetry breaking (SSB) are of
great importance in quantum condensed matter
physics~\cite{anderson}. Perhaps the most popular examples are
ferromagnetism and three-dimensional antiferromagnetism.
Off-diagonal long-range order (ODLRO) is the hallmark of
Bose-Einstein condensation and superconductivity~\cite{yang,kohn},
which may also be conveniently described in terms of SSB of gauge
symmetry. SSB of gauge symmetry is also important in high energy
physics.

In this Letter, we address the following question: what is the
feature of quantum  entanglement in the quantum states underlying
the phenomena of long-range order and spontaneous symmetry
breaking. As a special kind of correlation, quantum entanglement
refers to the situation that the quantum state of a composite
system is not a direct product of those of the
subsystems~\cite{epr}. It is an essential quantum
feature~\cite{bell,shi}. For many years, this concept has been of
central interest in foundations of quantum mechanics. Recently it
has been studied in the context of quantum information. As a basic
concept in quantum mechanics, it should  be useful for, and can be
studied in the context of,  many-body physics (cf.
Refs.~\cite{shi2,shi22,osterloh,osborn} and references therein).
Here we show that disentanglement, i.e. diminishing of
entanglement, underlies long-range orders, including both
off-diagonal and ``diagonal'' long-range orders. Ground state
disentanglement in presence of interaction provides a useful
insight on SSB and underlies the success of Landau theory of order
and phase transition, which is essentially classical. We also
explore entanglement characterizations suitable for many-body
physics.

\section*{2. Finite-temperature entanglement}

At zero temperature, a closed system is described as a pure state.
Hence one can use partial entropy $S(A)$ of a subsystem $A$ as the
measure of the bi-partite entanglement between $A$ and its
complementary subsystem~\cite{bennett1}.

At a finite temperature, a reasonable measure is the thermal
ensemble average of the entanglements in the Hamiltonian
eigenstates. For each Hamiltonian eigenstate $i$, as a pure state,
one can use partial entropy $S(A)$ of a subsystem $A$ as the
measure of the bi-partite entanglement  between  $A$ and its
complementary subsystem. Thus for a thermal ensemble, the
bi-partite entanglement is measured by $\langle S(A) \rangle
\equiv \sum_i p_i(T) S_i(A)$, where $p_i(T)$ is  the statistical
distribution at temperature $T$, $i$ denotes the Hamiltonian
eigenstates. The convexity  of entropy implies
\begin{equation}
S_{\rho(T)}(A) \geq  \langle S(A) \rangle \geq  0,
\end{equation}
where $\rho(T)=\sum_i p_i(T)|i\rangle\langle i|$, $S_{\rho(T)}(A)$
is the partial entropy of $A$ for $\rho(T)$.  Thus
$S_{\rho(T)}(A)$ is  the upper bound of the thermal average
entanglement $\langle S(A) \rangle$.

We remark that the ``mixed-state entanglement measures'' studied
in quantum information literature, e.g.  the so-called
concurrence~\cite{wootters}, are not suitable for thermal
ensembles in statistical physics.  In quantum information
literature, a ``mixed state entanglement''  is obtained by
considering all possible ensembles mathematically described by the
same density matrix. A thermal ensemble, on the other hand,  is
physically fixed in terms of the Hamiltonian eigenstates. With
this physical constraint, it is not physically  meaningful  to
decompose the thermal density matrix in terms of other ensembles.
For a density matrix of a subsystem obtained by tracing over the
complementary subsystem in a pure state of a larger system,
concurrence may be useful.

\section*{3. Density matrices and entanglement
of identical particles}

For a system of $N$  identical particles, consider the density
matrix  in Fock space,
$$\langle  n_1' \cdots  n_{\infty}|\rho|n_1 \cdots n_{\infty}\rangle,$$
where $n_k$ or $n_k'$  represent  the occupation number of the
single particle state $k$.  The physical constraints such as
particle number conservation and Pauli principle for fermions make
many matrix elements vanish.

The reduced density matrix  of the occupation-numbers of a set of
single particle states $1,\cdots,l$ is
$$\begin{array}{c}
\langle n_1'\cdots n_l'|\rho_l(1\cdots l)|n_1\cdots n_l\rangle \nonumber
\\
= \sum_{n_{l+1} \cdots n_{\infty}}
\langle n_1' \cdots n_l',n_{l+1} \cdots n_{\infty}|\rho|n_1 \cdots n_l,
n_{l+1} \cdots n_{\infty} \rangle. \nonumber \end{array}$$

For a pure state of a fixed number of identical particles, the
entanglement, in a given single particle basis,  means
superposition of different Slater determinants/permanents, and can
be quantified in terms of occupation-number
entanglement~\cite{shi22}.

The upper bound of the ensemble average entanglement between the
occupation-numbers of single particle states $1\cdots l$ and those
of other single particle states is thus given by the von Neumann
entropy of $\rho_l (1 \cdots l)$ obtained from $\rho(T)$. For
example, for one mode $k$, the von Neumann entropy of $\rho_1(k)$
is
\begin{equation}
S(k) = - \sum_{n_k}  \langle n_k| \rho_1(k) | n_k \rangle
log  \langle n_k| \rho_1(k) | n_k \rangle. \label{sk}
\end{equation}

For bosons, the summation in (\ref{sk}) is over $n_k=0,\cdots,N$,
$S(k)$ reaches the  maximal value $log N$  when $\langle n_k|
\rho_1(k) | n_k \rangle $ is the same for all these  values of
$n_k$. For fermions, the summation is over $n_k=0,1$, and the
maximum of $S(k)$ is  $log 2$. For both bosons and fermions,
$S(k)$ reaches the minimum $0$   when
 $\langle n_k| \rho_1(k) | n_k \rangle $ is $1$ for one value of
$n_k$ and is $0$ otherwise.  In general, the more inhomogeneous
the distribution of  $\langle n_k| \rho_1(k) | n_k \rangle $ for
different values of $n_k$, the smaller $S(k)$. Similar feature is
exhibited by the von Neumann entropy of the Fock-space reduced
density matrix of more than one mode, for the eigenmodes of this
reduced density matrix.

In terms of the states of the {\em  particles}, the density matrix
in {\em configuration space} is
$$\langle k_1'\cdots k_N'|\rho| k_1\cdots k_N \rangle,$$
while $i$-particle reduced density matrix is given by
%\begin{equation}
$$\langle k_1'\cdots k_i'|\rho^{(i)}|k_1\cdots k_i \rangle =
Tr (a_{k_1'}\cdots a_{k_i'} \rho a_{k_i}^{\dagger}\cdots a_{k_1}^{\dagger}),$$
%\end{equation}
with $Tr\rho^{(i)}=N(N-1)\cdots(N-i+1)$.

The following equation is a relation between the reduced density
matrices in configuration space and the reduced density matrices
in Fock space:
\begin{eqnarray}
Tr[\rho(a_k^{\dagger}a_k)^i] &=&\sum_{n_1 \cdots n_{\infty}}
n_{k}^i \langle n_1\cdots n_{\infty}|\rho|
n_1\cdots n_{\infty}\rangle \nonumber \\
&= &\sum_{n_k=1}^{N} n_k^i \langle n_k|\rho_1(k)|n_k\rangle,
\label{eq}
\end{eqnarray}
where $i=1,\cdots,N$.

We mention that for many-body systems,
the particle  reduced density
matrices are related to such quantities as density and distribution
functions, hence in principle the entanglement is experimentally
measurable.

\section*{4. ODLRO leads to disentanglement between the
condensate mode and the rest of system}

First let us consider bosons, for which one obtains from (\ref{eq}),
\begin{equation}
\left (
\begin{array}{llll}
1 & 2   & \cdots  & N   \\
1 & 2^2 & \cdots  & N^2 \\
\cdot& \cdot& \cdots &\cdot \\
1 & 2^N & \cdots  & N^N
\end{array}  \right)
\left(
\begin{array}{c}
 \langle 1|\rho_1(k)|1\rangle \\
 \langle 2|\rho_1(k)|2\rangle \\
 \cdots \\
 \langle N|\rho_1(k)|N\rangle
\end{array}
\right )
=
 \left(
\begin{array}{c}
 Tr(\rho a_k^{\dagger}a_k) \\
 Tr[\rho(a_k^{\dagger}a_k)^2] \\
 \cdots \\
Tr[\rho(a_k^{\dagger}a_k)^N]
\end{array}
\right ).  \label{a}
\end{equation}

Bose-Einstein condensation is characterized by ODLRO in one-particle
reduced density matrix
$\rho^{(1)}$, i.e.
%\begin{equation}
$\langle {\mathbf{x}}'|\rho^{(1)}|{\mathbf{x}} \rangle  \neq  0 \qquad
 \mbox{as} \qquad  |{\mathbf{x}}-{\mathbf{x}}'| \rightarrow \infty$,
%\end{equation}
which  is  equivalent to the existence of an eigenvalue of order
$N$, i.e
\begin{equation}
\rho^{(1)}  = \lambda_0^{(1)} |\lambda_0^{(1)}\rangle\langle
\lambda_0^{(1)}|+ \sum_{j\neq 0}
\lambda_j^{(1)}|\lambda_j^{1}\rangle\langle \lambda_j^{(1)}|,
\label{def}
\end{equation}
where $\lambda_0^{(1)}=N\alpha$, $\alpha$ is a
finite fraction. Hence (\ref{def}) can also be
used as a  definition of Bose-Einstein condensation~\cite{leggett}.

Let us  consider Eq.~(\ref{a}) in the eigen-basis of $\rho^{(1)}$,
i.e. $\{|k\rangle\}$ is given by  $\{|\lambda^{(1)}_j\rangle\}$.
Let $|k_0\rangle \equiv |\lambda_0^{(1)} \rangle$. For a given set
of $Tr[\rho(a_{k}^{\dagger}a_{k})^i]$, $i=1,\cdots,N$, there is a
unique  set of $\langle n_{k}|\rho_1(k)|n_{k}\rangle$, where
$n_k=0,1,\cdots,N$. For  $k=k_0$, $Tr(\rho
{a_{k_0}^{\dagger}}{a_{k_0}}) = N\alpha$, while
$Tr[\rho(a_{k_0}^{\dagger}a_{k_0})^i] \approx  N^i\alpha^i$ for
$i=2,\cdots, N$.   If $Tr[\rho(a_{k_0}^{\dagger}a_{k_0})^i]$ is
exactly $N^i\alpha^i$ and $N\alpha$ is an integer, e.g. when
$\alpha=1$, then $\langle N\alpha|\rho_1(k_0)|N\alpha\rangle=1$
while $\langle n_{k_0}|\rho_1(k_0)|n_{k_0}\rangle=0$ for
$n_{k_0}\neq N\alpha$, consequently $S(k_0)=0$. More generally,
the values of $\langle n_{k_0}|\rho_1(k_0)|n_{k_0}\rangle$
corresponding to one or very few values of  $n_{k_0}$ very close
to $N\alpha$ are finite fractions. Since $\sum_{n_{k_0}} \langle
n_{k_0}|\rho_1(k_0)|n_{k_0}\rangle=1$, there can be only  very
few, typically only one, finite  fraction.

Thus ODLRO at $k_0$ generally implies that  $S(k_0)$ is very
small. Typically, suppose  $\langle
I(N\alpha)|\rho_1(k_0)|I(N\alpha)\rangle$ is  a finite fraction
$\gamma$, where  $I(N\alpha)$  denotes the  integer closest to
$N\alpha$, while $\langle n_{k_0}|\rho_1(k_0)|n_{k_0}\rangle$ for
$n_{k_0} \neq I(N\alpha)$ is of the order of  $(1-\gamma)/N$. Then
$S(k_0) \approx  -\gamma \log \gamma$, which is very small. With
$S(k)$ being the upper bound, the bi-partite entanglement between
the occupation-number  at $k_0$  and the rest of the system is
thus also very small, approaching $0$ when $\alpha \rightarrow 1$.
We refer to such diminishing of entanglement as disentanglement.

Therefore  Bose-Einstein condensation signals disentanglement
between the occupation-number of the condensate mode and the  rest
of the  system. Likewise, for a fragment condensation, in which
there are more than one condensate mode,  disentanglement occurs
respectively between each condensate mode and its complementary
subsystem.

Now consider fermions. From Eq.~(\ref{eq}), one obtains $Tr[\rho
(a_k^{\dagger}a_k)^i] = \langle 1|\rho_1(k)|1\rangle \leq 1$,
which does not lead to a particular  specification on the nature
of entanglement between one mode and others. Indeed, there cannot
be ODLRO in $\rho^{(1)}$~\cite{yang},  thus there is no
ODLRO-induced disentanglement between  one fermion mode and others
(the non-entangled ground state of a fermi liquid~\cite{shi2} is
not the concern here). Moreover, one can evaluate $\langle n_{k_1}
n_{k_2}\rangle$, $\langle (n_{k_1}+1)n_{k_2}\rangle$, $\langle
n_{k_1}(n_{k_2}+1)\rangle$ and $\langle
(n_{k_1}+1)(n_{k_2}+1)\rangle$, obtaining $\langle
00|\rho_2(k_1,k_2)|00\rangle = \langle n_{k_1}n_{k_2}\rangle-
\langle n_{k_1}\rangle-\langle n_{k_2}\rangle+1$, $\langle
01|\rho_2(k_1,k_2)|01\rangle = \langle n_{k_2}\rangle- \langle
n_{k_1}n_{k_2} \rangle$, $\langle 10|\rho_2(k_1,k_2)|10\rangle =
\langle n_{k_1}\rangle - \langle n_{k_1}n_{k_2}\rangle$, $\langle
11|\rho_2(k_1,k_2)|11\rangle = \langle n_{k_1}n_{k_2}\rangle$. It
can be seen that there is no ODLRO-induced disentanglement between
the occupation-numbers of two fermion  modes   and the rest of the
system.

However, for both bosons and fermions,  there can be ODLRO in the
two-particle reduced density matrix, i.e.
\begin{equation}
\rho^{(2)}  = \lambda_0^{(2)} |\lambda_0^{(2)}\rangle\langle
\lambda_0^{(2)}|+ \sum_{j \neq 0}
\lambda_j^{(2)}|\lambda_j^{(2)}\rangle\langle \lambda_i^{(2)}|,
\label{def2}
\end{equation}
where $\lambda_0^{(2)}=N\delta$, $\delta$ is a finite fraction.
$\sum_j \lambda_j^{(2)}= N(N-1)$.  For fermions $\lambda_0^{(2)}
\leq N$, and  ODLRO in $\rho^{(2)}$ is a characterization of
superconductivity~\cite{yang}. Note the difference between ``a
two-particle mode'' and  ``two one-particle modes''. The former is
a unitary transformation of the latter  in the two-particle
Hilbert space, and can be written as $|K \rangle = \sum_{k_1,k_2}
U_{K,(k_1,k_2)}|k_1,k_2\rangle$. The associated creation operator
is $b_K^{\dagger} = \sum_{k_1,k_2} U_{K,(k_1,k_2)}
a_{k_1}^{\dagger} a_{k_2}^{\dagger}$. $b_K^{\dagger}b_K$ gives the
number of  particle pairs in mode $K$, with the maximum $N(N-1)$.
There are overlaps between different pairs in terms of the
original particles.  One obtains
\begin{equation}
Tr[\rho(b_K^{\dagger}b_K)^j] = \sum_{n_K=1}^{N(N-1)} n_K^j \langle
n_K|\rho_2(K)|n_K\rangle,
\label{r2}
\end{equation}
with $j=1,\cdots,N(N-1)$.

Following an argument similar to the above  one for ODLRO in
$\rho^{(1)}$, one can find  that ODLRO in $\rho^{(2)}$ implies
disentanglement  between the  occupation-number  of the
two-particle condensate mode $|\lambda_0^{(2)}\rangle$ and the
rest of the system.

In general, for both bosons and fermions, it can be shown that if
there is ODLRO in $\rho^{(i)}$, there is disentanglement between
the occupation-number of the  $i$-particle condensate mode and
others in the eigen-basis of $\rho^{(i)}$. For bosons, ODLRO in
$\rho^{(i)}$ implies ODLRO in $\rho^{(j)}$ with $j >
i$~\cite{yang} and thus also disentanglement between  the
occupation-number of the  $j$-particle condensate  mode and its
complementary subsystem.

The disentanglement of the occupation number of the condensate
mode from the system is consistent with the well known result, as
used in Bogoliubov theory, that the occupation number of the
condensate mode is approximately a constant, which implicates that
the system is an eigenstate of the occupation number of the
condensate mode. This disentanglement also justifies the
(classical) two-fluid model of superfluidity.

\section*{5.
Long-range order and spontaneous Symmetry breaking}

Disentanglement also underlies ``diagonal'' long-range orders,
e.g.   ferromagnetic  state $|\uparrow\dots\uparrow\rangle$ and
antiferromagnetic N\'{e}el state
$|\uparrow\downarrow\uparrow\dots\uparrow\downarrow\rangle$, which
are product states. They are enforced by energetics and
spontaneous symmetry breaking (SSB). Suppose the square of the sum
of the spin operators is $S^2$, which commutes the Heisenberg
Hamiltonian.  The lowest energy  states of a ferromagnet is the
eigenstates with $S^2=Ns(Ns+1)$, where $N$ is the number of sites,
$s$ is  each spin. The lowest energy state of an antiferromagnet
is  a singlet $S^2=0$. Because of SSB,  the physical  ground state
of a ferromagnet is a ferromagnetic state in a certain direction,
rather than a superposition state of ferromagnetic states in
different directions. The antiferromagnetically ordered state,
i.e. a  N\'{e}el state, is  not even an energy eigenstate.

The point of view of disentanglement can provide insights on the
nature of SSB. For large $N$, a superposition state is a
macroscopic superposition, which is  highly fragile under
perturbations. As a ``Schr\"{o}dinger cat'', it  normally reduces
to a basis product state. The ferromagnetic state or
antiferromagnetic N\'{e}el state is favored over other basis
states because they correspond to the lowest energy among the
basis states. Ferromagetism and antiferromagnetism represent  two
different types of SSB~\cite{peierls}. Disentanglement appears to
provide a unified insight. Usually SSB is attributed to the near
degeneracy between the symmetric states and thus the stability of
the symmetry-breaking states. Complementarily, decoherence due to
the coupling with the environment is also useful in explaining
SSB. More discussions on this aspect will be made elsewhere.

On the other hand, the stabilization of the singlet  state in  a
low dimensional antiferromagnet  may be understood as due to
higher tunnelling rate between different product basis  states, or
lower decoherent rate of the singlet state. This point of view may
be supported by the fact that both the tunneling rate and the
scattering cross section have dimensional dependence (in general,
decoherence rate may be proportional to a certain scattering cross
section, cf.~\cite{dec}). Without perturbation, the tunnelling
between different ferromagnetic states is zero in any dimension
since it is a Hamiltonian eigenstate.

Disentanglement also provides insights on SSB of gauge symmetry,
as a description  of Bose condensation and superconductivity.  For
a closed system, this description is an
approximation~\cite{peierls}. We think that the excellence of this
approximation  is not only because of giving the peaked particle
number and energy, but also because ODLRO or disentanglement makes
it a good approximation to write $\langle
\hat{\psi}^{\dagger}({\mathbf{x}}') \psi({\mathbf{x}})\rangle$ as
$\langle \hat{\psi}^{\dagger}({\mathbf{x}}')\rangle\langle
\psi({\mathbf{x}})\rangle$, where the average is over a particle
number non-conserved (``coherent'') state.  The difference with
the genuine SSB is that it is merely determined by energetics,
disentanglement happens without external perturbation or
environment-induced decoherence. For an open system, it   may be
viewed as a genuine SSB~\cite{anderson,leggett2}. It may be
understood as that,  like antiferromagnetism, the system
disentangles or decoheres into a ``coherent state'', which is not
the Hamiltonian eigenstate. The reason why the ``coherent state''
basis is favored may be related to its robustness~\cite{zurek}.

With the long-range order,  the system is characterized by the
order parameter, which is usually given  by the average
expectation value of the concerned operator, e.g. $\langle \sum
{\hat{s}_{iz}} \rangle$ for the ferromagnetism or $\langle
\hat{\psi}({\mathbf{x}}) \rangle $ for the Bose condensation. With
disentanglement, it is directly related to the state of each
single particle. In fact, the order parameter of Bose-Einstein
condensation  can be directly chosen to be the single particle
wavefunction~\cite{leggett}. We see that just because the quantum
state is, to the zeroth-order approximation, a (disentangled)
product of a  same single particle state (in the case of N\'{e}el
state, it is a product of two opposite spin states),  it can be
described by such an  order parameter, upon which the  Landau
theory is based. The quantum fluctuation over the order parameter,
for example, the spin wave in a ferromagnet or the quantum
correction in Bose-Einstein condensation, is related to the small
nonzero entanglement.

In relativistic quantum field theory, with SSB, the scaler field
is in a particular vacuum among the degenerate vacua, rather than
a superposition of different vacua. Similar to the condensed
matter cases, usually the SSB is assured by the vanishing of the
matrix elements between symmetry breaking vacua. We leave for
future discussions the subtle details related to the present
discussion, for example, whether decoherence due to the coupling
with another degree of freedom, say, the gauge field, may be
possible, and the nature of entanglement in the symmetry breaking
vacuum.

\section*{6. Correlation functions and fluctuations}

Several quantities are used in characterizing  order or
fluctuation. The first is  correlation function (or staggered
correlation function in the case of antiferromagnetism), i.e. the
average of products of operators at  different sites, e.g.
$\langle \hat{s}_{iz}\hat{s}_{jz}\rangle$. The second is the
connected correlation function, e.g. $\langle
\hat{s}_{iz}\hat{s}_{jz}\rangle_c \equiv \langle \hat{s}_i
\hat{s}_j\rangle-\langle \hat{s}_i\rangle \langle
\hat{s}_j\rangle$. The third is the fluctuation amplitude of an
operator $\hat{O}$, given by $\langle \hat{O}^2 \rangle -\langle
\hat{O} \rangle^2$.

Long-range order means the nonvanishing of, say, $\langle
\hat{s}_{iz} \hat{s}_{jz}\rangle$ when the distance between $i$
and $j$ approaches infinity.  It reaches maximum when the state is
the ferromagnetic or N\'{e}el state. In a generic  superposition
state, it is small for large distance. It would  still be large if
the state could be a superposition of different ferromagnetic or
antiferromagnetic N\'{e}el state, which is, however, excluded by
SSB.

Thus long-range order in $\hat{s}_{iz}$, as quantified by the
correlation function, may signal  the disentanglement into the
ferromagnetic or antiferromagnetic state in $z$ direction.

The relation between disentanglement and long-range order can also
be exemplified by the two limits of a quantum Ising model in a
transverse field, $H=J\sum \hat{s}_{iz}\hat{s}_{jz}+B\sum
\hat{s}_{ix}$. In the strong interaction limit, as a simple Ising
model, there is SSB, long-range order and disentanglement, as
discussed above. In the strong external coupling limit, the state
is also disentangled. While there is no long-range order in
$\hat{s}_{iz}$, there is long-range order in $\hat{s}_{ix}$,
though it is due to the external coupling. With both coupling and
interaction nonzero, any eigenstate of this Hamiltonian is always
entangled~\cite{shi2}. Roughly speaking, entanglement at zero
temperature, as an alternative to thermal fluctuation, provides
the nonvanishing connected correlation function, as detailed in
the next section. Some behavior of entanglement was recently
investigated in a one-dimensional model~\cite{osterloh,osborn}.

In low dimensional antiferromagnets,  spin liquid states, e.g. the
RVB state, become important. This  may be understood as due to the
diminishing of the  effect of SSB or disentanglement. They are
indeed entangled states. The amount of entanglement for a
short-range RVB state on a square lattice~\cite{kivelson} has been
calculated~\cite{shi2}. On the other hand, it has been known that
the staggered correlation function in this state is exponentially
bounded~\cite{kohmoto}. This is a converse example of our argument
concerning the relation between long-range order and
disentanglement.

In a pure quantum state, the fluctuation amplitude is nonzero  if
and only if the state is not an eigenstate of the operator, i.e.
the state is a superposition of its eigenstates. So if the
operator involves more than one particle, e.g. if $\hat{O}=\sum_i
\hat{s}_{iz}$, there may be entanglement. This can be seen from
$\langle \hat{O}^2\rangle - \langle \hat{O}\rangle^2 = \sum_i
(\langle \hat{s}_{iz}^2\rangle - \langle \hat{s}_{iz} \rangle ^2)
+\sum_{i\neq j} \langle \hat{s}_{iz}\hat{s}_{jz}\rangle_c$. It
will be shown below that a nonzero connected correlation function
$\langle \hat{s}_{iz}\hat{s}_{jz}\rangle_c$ can characterize the
entanglement between any two parts with $i$ and $j$ belonging to
them respectively. Therefore {\em the quantum fluctuation of a sum
of local operators is the sum of the quantum fluctuations of
individual local operators and all the two-body entanglements}.

At a finite  temperature, presumably the fluctuation contains both
thermal and quantum ones. However, if the Hamiltonian is
compatible with the operator in question, then the fluctuation  is
solely thermal. Moreover, although in general a connected
correlation function is contributed by both thermal fluctuations
and  quantum entanglement,  with disentanglement, it is only  due
to thermal fluctuation. Thermodynamic entropy only measures the
population of Hamiltonian  eigenstates, therefore thermal phase
transitions,  determined by the competition between entropy and
energetics  and  associated with the change of  order and
symmetry, is essentially classical (although the average
entanglement may change with the temperature simply because
different Hamiltonian eigenstates may contain different amounts of
entanglement).

\section*{7.  Connected correlation functions
as characterizations of entanglements in  a pure state}

In this section, we discuss how the connected correlation
functions can be used in characterizing  entanglement. This
approach has the advantage that it is not necessary to explicitly
know the many-body state, and thus quite suits many-body systems.
It is naturally connected with the traditional many-body
techniques, and can be used to study, for example, the pairwise
entanglement as a function of the distance.

For a pure state $|\psi\rangle$, the connected correlation
function of two operators, for two bodies $i$ and $j$
respectively, $\langle \hat{O}_{i}\hat{O}_{j}\rangle_c\equiv
\langle \hat{O}_{i}\hat{O}_{j}\rangle-\langle \hat{O}_{i}\rangle
\langle \hat{O}_{j}\rangle$, vanishes if $|\psi\rangle$ is any
direct product of two factors to which $i$ and $j$ belong
respectively~\cite{ex}.

Therefore if for certain operators $\hat{O}_{i}$ and
$\hat{O}_{j}$, $\langle \hat{O}_{i}\hat{O}_{j}\rangle_c \neq 0$,
then there is not any bi-partition of the system, with $i$ and $j$
belonging to the two different parts, such that $|\psi\rangle$ is
a direct product of the pure states of the two parts, i.e. any
part of the system with $i$ included is entangled with its
complementary part with $j$ included.

In general, the connected correlation function  of the operators
of $n$ bodies  is  the correlation function of these $n$ operators
deducted by all kinds of products of the connected correlation
functions of proper subsets of  these $n$ operators. It measures
the part of the correlation which is not due to the correlations
of not all of the bodies. For example,
\begin{equation}
\begin{array}{lll}
\langle \hat{O}_{i}\hat{O}_{j}\hat{O}_{k}\rangle_c &\equiv&
\langle
\hat{O}_{i}\hat{O}_{j}\hat{O}_{k}\rangle- \\
&&(\langle \hat{O}_{i}\rangle \langle \hat{O}_{j} \rangle
\langle \hat{O}_{k}\rangle + \\
&&\langle \hat{O}_{i} \hat{O}_{j} \rangle_c \langle
\hat{O}_{k}\rangle + \langle \hat{O}_{i} \hat{O}_{k} \rangle_c
\langle \hat{O}_{j}\rangle + \langle \hat{O}_{j} \hat{O}_{k}
\rangle_c \langle \hat{O}_{i}\rangle),
\end{array}
\end{equation}

\begin{equation}
\begin{array}{lll}
\langle \hat{O}_{i}\hat{O}_{j}\hat{O}_{k}\hat{O}_{l}\rangle_c
&\equiv& \langle
\hat{O}_{i}\hat{O}_{j}\hat{O}_{k}\hat{O}_{l}\rangle- \\
&&(\langle \hat{O}_{i}\rangle \langle \hat{O}_{j} \rangle
\langle \hat{O}_{k}\rangle \langle \hat{O}_{l}\rangle  \\
&&+\langle \hat{O}_{i} \hat{O}_{j} \rangle_c \langle
\hat{O}_{k}\rangle \langle \hat{O}_{l}\rangle + \cdots (6 \mbox{ }
terms)
\\
&&+\langle \hat{O}_{i} \hat{O}_{j} \rangle_c \langle \hat{O}_{k}
\hat{O}_{l}\rangle_c + \cdots (3 \mbox{ } terms)\\
&&+\langle \hat{O}_{i} \hat{O}_{j}\hat{O}_{k} \rangle_c \langle
\hat{O}_{l}\rangle + \cdots (4 \mbox{ } terms)), \\
\end{array}
\end{equation}
and so on.

It can be seen that if $|\psi\rangle$  {\em contains} a separable
factor of a subset of  $m$ parts, and suppose $i_1$, $i_2$,
$\cdots$, $i_m$ are $m$ bodies belonging to these $m$ parts
respectively, then any connected correlation function of these $m$
bodies vanishes. Of course, similar is for any connected
correlation function of a subset of these $m$ bodies. For example,
suppose
$|\psi\rangle=|\phi(ij\cdots)\rangle\otimes|\phi(k\cdots)\rangle$,
where $\cdots$ denotes the bodies other than $i$, $j$ ,$k$ if
there are other bodies in the system. Then
$\langle\hat{O}_{i}\hat{O}_{k}\rangle_c=
\langle\hat{O}_{j}\hat{O}_k\rangle_c=0$. Whether
$\langle\hat{O}_{i}\hat{O}_{j}\rangle_c$  vanishes depends on
whether $i$ and $j$ can further be separated. In either case, it
is certain that $\langle
\hat{O}_{i}\hat{O}_{j}\hat{O}_{k}\rangle_c =0$. If
$|\psi\rangle=|\phi(i\cdots)\rangle
\otimes|\phi(j\cdots)\rangle\otimes|\phi(k\cdots)\rangle$, then
$\langle\hat{O}_{i}\hat{O}_{k}\rangle_c=
\langle\hat{O}_{j}\hat{O}_k\rangle_c=
\langle\hat{O}_{i}\hat{O}_{j}\rangle_c= \langle
\hat{O}_{i}\hat{O}_{j}\hat{O}_{k}\rangle_c =0$.

Therefore if for certain operators $\hat{O}_{i_1}$, $\cdots$,
$\hat{O}_{i_m}$, $\langle
\hat{O}_{i_1}\cdots\hat{O}_{i_m}\rangle_c \neq 0$, then for any
$m$-partite partition of the system, with $i_1$, $\cdots$, $i_m$
belonging to the different $m$ parts, $|\psi\rangle$ does not {\em
contain any} separable factor of  any proper subset of the $m$
parts. That is to say, there exists {\em true $m$-partite
entanglement} among these $m$ parts, i.e. the entanglement cannot
be reduced to the entanglement among a proper subset of the $m$
parts. Thus a non-vanishing  $m$-body connected correlation
function characterizes the true $m$-partite entanglement.  For
example, if $\langle \hat{O}_{i}\hat{O}_{j}\hat{O}_{k}\rangle_c
\neq 0$, then $|\psi\rangle$ cannot be separated as
$|\phi(i\cdots)\rangle\otimes|\phi(j\cdots)\rangle\otimes|\phi(k\cdots)\rangle$
or $|\phi(ij\cdots)\rangle\otimes|\phi(k\cdots)\rangle$ or
$|\phi(ik\cdots)\rangle\otimes|\phi(j\cdots)\rangle$ or
$|\phi(jk\cdots)\rangle\otimes|\phi(i\cdots)\rangle$.

Now we consider the examples of disentanglement studied in the
previous sections. The ferromagnetic or antiferromagnetic N\'{e}el
state is completely separable, i.e. it is a product of the pure
spin states of all the spins, hence any connected correlation of
any set of spins vanishes. For ODLRO, the above discussion implies
that at zero temperature, the occupation number of condensate mode
is separated from the rest of the system, therefore any connected
correlation function of the occupation numbers of any number of
modes with one of which being the condensate mode  must vanish.

What about the converse, i.e.  when does the  vanishing of a
connected correlation imply separability? In the following, we
only  consider the simplest case: a two-spin state. Without loss
of essence, by considering  Schmidt decomposition,  the state can
be written as
\begin{equation}
|\psi\rangle=\cos\theta|\alpha_i\rangle|\gamma_j\rangle
+\sin\theta|\beta_i\rangle|\delta_j\rangle, \label{as}
\end{equation}
 where $|\alpha_i\rangle$ and $|\beta_i\rangle$, as eigenfunctions
 of the spin operator $\hat{s}_{\mathbf{n}_i}$ in a certain direction, comprise a spin
 basis of $i$,  while $|\gamma_j\rangle$ and $|\delta_j\rangle$,
  as eigenfunctions of
 a certain $\hat{s}_{\mathbf{n}_j}$,  comprise a spin
 basis of $j$.  The basis states for the two-spin system
 can be chosen to be
$|\chi_1\rangle=|\psi\rangle$,
$|\chi_2\rangle=-\sin\theta|\alpha_i\rangle|\gamma_j\rangle
+\cos\theta|\beta_i\rangle|\delta_j\rangle$,
$|\chi_3\rangle=|\alpha_i\rangle|\delta_j\rangle$ and
$|\chi_4\rangle  = |\beta_i\rangle|\gamma_j\rangle$. Then
$\langle\hat{s}_{\mathbf{n}_i}\hat{s}_{\mathbf{n}_j}\rangle_c
=\langle\psi|\hat{s}_{\mathbf{n}_i}\hat{P}^{\perp}\hat{s}_{\mathbf{n}_j}|\psi\rangle$.
 where $\hat{P}^{\perp}\equiv 1- |\psi\rangle\langle\psi|=
|\chi_2\rangle\langle\chi_2|+|\chi_3\rangle\langle\chi_3|+
|\chi_4\rangle\langle\chi_4|$ is the projection onto the subspace
orthogonal to $|\psi\rangle$. With eigenvalues of each spin
operator being $\pm 1/2$, one obtains $|\langle
\hat{s}_{\mathbf{n}_i}\hat{s}_{\mathbf{n}_j}\rangle_c| =
\sin^2\theta\cos^2\theta$.

Therefore if $\langle
\hat{s}_{\mathbf{n}_i}\hat{s}_{\mathbf{n}_j}\rangle_c =0$, then
$|\psi\rangle$ is a direct product of pure states of $i$ and $j$.
To use this result, one first needs to find
$\hat{s}_{\mathbf{n}_i}$ and $\hat{s}_{\mathbf{n}_j}$  by, say,
diagonalizing the reduced density matrices of each spin.

Finally we mention that all the connected correlation functions of
operators $\hat{O}_1,\hat{O}_2,\cdots$ can be obtained from a
generating functional $F\{\mathbf{h}\} \equiv \ln
Z\{\mathbf{h}\}$, with $Z\{\mathbf{h}\}=\langle
e^{\mathbf{h}\cdot\mathbf{\hat{O}}}\rangle$, $\mathbf{h}\equiv
(h_1,h_2,\cdots)$, $\mathbf{\hat{O}}\equiv
(\hat{O}_1,\hat{O}_2,\cdots)$, where the subscripts denote the
different bodies in the system. The connected correlation
functions of $\hat{O}_{i_1}$, $\cdots$, $\hat{O}_{i_n}$ can be
obtained as
\begin{equation}
\langle O_{i_1}\cdots\hat{O}_{i_n}\rangle_c =
\frac{\delta^nF}{\delta h_{i_1}\cdots\delta
h_{i_n}}\mid_{\mathbf{h}=0}.
\end{equation}

\section*{8. Summary}

We have shown that off-diagonal long-range order leads to
disentanglement between the condensate mode and the rest of the
system. Furthermore, it is revealed that in general, diminishing
of entanglement underlies various long-range orders and
spontaneous symmetry breaking. This is consistent with the wisdom
that Landau theory of order and symmetry breaking is essentially
classical, even though the order parameter has a quantum
origin~\cite{wen}. Remarks are also made on the relations between
entanglement on one hand, and fluctuation and correlation
functions on the other. Entanglements in a pure state can be
characterized in terms of the nonvanishing  connected correlation
functions.

\section*{Acknowledgements}

This work was partly supported by the program grant of TCM group
of Cavendish Laboratory and was also an output from project
activity funded by The Cambridge-MIT Institute Limited.


\begin{thebibliography}{99}
\bibitem{ex} The special case that
a two-body connected correlation function vanishes for a product
state was also noted in Ref.~\cite{osborn} (under the term
``spin-spin correlation function'').
\end{thebibliography}
\end{document}